\documentclass[aps,apl,preprint]{revtex4-1}
\usepackage{amsmath}
\usepackage{amssymb}
\usepackage{graphicx}
\usepackage{times}
\usepackage{hyperref}
\usepackage{float}

\mathchardef\mhyphen="2D

\begin{document}



\title{Ultrafast control of  inelastic tunneling in a double semiconductor quantum well}
%
%
\author{Michael Sch\"uler, Yaroslav Pavlyukh, Jamal Berakdar}
\affiliation{Institut f\"ur Physik, Martin-Luther-Universit\"at Halle-Wittenberg, Von-Seckendorff-Platz 1,
06120 Halle, Germany}
%

\date{\today}

\begin{abstract}
In a semiconductor-based double quantum well (QW) coupled to a degree of freedom with an
internal dynamics, we demonstrate that the electronic motion is controllable within
femtoseconds by applying appropriately shaped electromagnetic pulses.  In particular, we
consider a pulse-driven Al$_x$Ga$_{1\mhyphen x}$As based symmetric double QW coupled to
uniformly distributed or localized vibrational modes and present analytical results for
 the lowest two levels. These predictions are
assessed and generalized by full-fledged numerical simulations showing that localization
and time-stabilization of the driven electron dynamics is indeed possible under the
conditions identified here, even with a simultaneous excitations of vibrational modes.
\end{abstract}
\maketitle The impressive advance in fabrication, characterization and control of
nano-scale systems \cite{handbook} has fueled wide-ranged studies on the fundamental
behavior of quantum systems and their utilization for practical applications, e.g. as
elements in nano-circuits \cite{nanocirc}, nano-oscillators \cite{nanoosc}, nano-magnets
\cite{nanomag} and opto-electronic devices \cite{handbook}. A similar rapid progress was
also achieved in optics. Light sources are currently available in a wide range of
intensity, duration and pulse shapes \cite{handbook_optics}, offering new possibilities
for the temporal and local manipulations of electronic properties of nanostructures.  A
paradigm example has been the ultrafast control of quantum states in a double-well (DW)
structure which is of relevance for quantum computing application \cite{dwp}.  In this
respect, three aspects are important: i) the swift electron localization in one of the
wells starting from an arbitrary initial state, ii) maintaining this localization for a
desired time, and iii) switching of the localization between the two wells.  From a
practical point of view, it is also essential to address the role of the coupling to the
environment.  In this work we study the electron dynamics in a typical
Al$_x$Ga$_{1\mhyphen x}$As-based QW coupled to vibrational modes with the aim of
controlling and sustaining the electron motion on a femtosecond time scale via
electromagnetic pulses.
The DW potential barrier $V_{conf}(x) $ with a well separation $x_0$ confines the
conduction electrons in the $x$-direction and is well modelled by $V_{conf}(x) = 16~
(V_B/x_0^4) \left( x^2 - x_0^2/4 \right)^2$.  Typical values for this system are
$V_B \approx 240$~meV and $x_0 \approx 10$~nm. The effective mass is assumed constant
across the structure $m \approx 0.066 m_e$ (as for GaAs, and $m_e$ is the free electron
mass).
The tunnel splitting $\hbar \Omega = E_2-E_1$ where $E_1$ and $E_2$ are the two lowest
eigenenergies, is smaller than the separation from the third excited state, hence a
consideration within a two level system (TLS) approximation is useful for low excitations
and will be employed below for the analysis of qualitative trends.  In addition we couple
the electron dynamics to a degree of freedom ($y$) with an internal vibrational dynamics
with the Hamiltonian $H_R = p_y^2/2m + k y^2/2$ and a coupling potential $H_C$. The full
Hamiltonian reads then $H = H_0 + H_C + H_R$. Here $H_0$ is the bare electron term; the
coupling has the form $H_C = V_0 (y/y_0) \ v_c(x)$ with the function $v_c(x)$ describing
the interaction strength along the structure, $y_0$ stands for the typical length scale
associated with the oscillator potential. This problem resembles the
rotational dynamics coupled to vibrations \cite{volker} and also inelastic scanning tunneling experiments \cite{bringer}.
 We will discuss two cases: (i) a uniform
harmonic force acting on the electron, i.e. the coupling potential is $v_c(x) = x/x_0$,
and (ii) a localized coupling $v_c(x) = \exp[-(x-x_c)^2/2 \lambda^2]$.  Physically such
vibrational (phononic) coupling may arise due to inelastic scattering of the tunneling
electron in the DW, e.g. from pinned impurities that then vibrate around the minimum of
the pinning potential.

A particularly efficient and swift mean for steering the electron dynamics in DW is the
use of specially shaped femtosecond light pulses such as the so-called half cycle pulses
(HCPs) \cite{hcp,matoslocal}.  These are strongly asymmetric pulses with one short and
strong half-cycle followed by a very weak and long half-cycle of opposite polarity.  A key
point in this control scheme is that the pulse duration has to be much shorter than the
characteristic time of DW system ($T_0 = 2 \pi / \Omega$) \cite{matoslocal}. In this case,
within the dipole approximation, the action of the pulses can be viewed effectively as
impulsive kicks \cite{kick} delivered at times $t_k$ with a strength $\Delta \mathbf{p}_k
=\int \tilde{\mathbf{E}}_k dt $ where $\tilde{\mathbf{E}}_k$ is the electric field
amplitude of the pulse and $t_k$ is the time when $\tilde{\mathbf{E}}_k$ reaches its peak,
meaning that such a pulse sequence acts in effect as $ {\mathbf{E}}_k = \Delta
\mathbf{p}_k \delta(t-t_k) $ and couples to the electron via $H_L(t) = x
E(t)$~\cite{remark}. This type of pulses can localize the electron within femtoseconds
whereas the localization time for harmonic pulses is about few picoseconds \cite{bavli}.
How effective this scheme would remain if the tunneling electrons are coupled to an
external degree of freedom is to be clarified here.  Physically one would expect a
similarly effective control scheme when the electron dynamics is coupled to an external
bath as long as the time scales involved are longer than the pulse duration. However, the
whole process is a coupled dynamical one with a non-trivial dependence on the properties
of the external degree of freedom.  We start thus from the two-level approximation. The
case of zero vibrational coupling has been analyzed analytically
\cite{matoslocal,matosdoc}. The electron in the ground state is completely delocalized
across the heterostructure.  In what follows we assume that in the ground state the
electrons are decoupled from the vibrations.  The vibrational coupling acts on the excited
states, i.e. as soon as the external pulses are applied.  Starting from this state, in
order to localize the electron within the left well, it is sufficient to apply a single
$\pi/2$-pulse in $x$-direction, i.e. $E(t) = \Delta p \ \delta(t) $ with $ \alpha = 2 \mu \Delta p/\hbar = \pi/2$,
where $\mu$ being the transition dipole between the first two states and $\alpha$ can be viewed as
 the "pulse angle". After acquiring an additional momentum shift $\Delta p$ the electron
tunnels to the left well within the time $\tau_L = T_0/4$. As the system undergoes free
Rabi-oscillations, the probability of finding the electron in the left well $P_L$ is given
by $P_L(t) = \cos^2(\Omega t/2)$. Within the half-period of the Rabi oscillation
the electron  is then mainly localized in the right quantum well.
To maintain the localization, one may  kick periodically the electron back by an
appropriate laser pulse. Applying the $\pi$-pulse ($\alpha  =
\pi$) to the free-evolving system at $t=t_0$ brings the system back to the localized state
at $t=2t_0$ so that $P_L(0) = P_L(2 t_0) = 1$. Therefore, the system has the periodicity
$T$ of the array of laser pulses with $T = 2 t_0$. It is clear that the averaged
localization will be higher if the interval $T$ between two consecutive pulses is shorter.

To proceed  further, we map the electron Hamiltonian $H_0$ onto the spin-like TLS
Hamiltonian $H_{TLS} = - \hbar \Omega \sigma_z / 2$ (with $\sigma_{j}$ being the $j$ Pauli matrix).
 The mapping can be extended to
incorporate the vibrational coupling by substituting the position operator with $-x_0
\sigma_x/2$ (with eigenstates localized in the left, right potential minima). Applying
this rule to the coupling term $v_c(x)$ leads to the so called spin-boson-model (see
\cite{weiss} and ref. therein). According to the theory of small polarons
\cite{gitterman}, the remaining vibrational degree of freedom $y$ is incorporated as a
renormalization of the tunneling amplitude $\Omega \rightarrow \tilde \Omega = M_{n,n}
\Omega$.
Thus, the polaron
transformation decouples the electron from the vibrational degrees of freedom, so that the
remaining renormalized electron Hamiltonian can be handled analytically. We will use this
model to make predictions for the modified control mechanism and compare with the full
numerical exact result.
\paragraph*{(i) Uniform coupling $v_c(x) = x/x_0$.} The effective electron Hamiltonian
reduces to the simple form $\tilde H_{TLS} = - \hbar \tilde \Omega \sigma_z/2$ with the
renormalized tunneling amplitude $\tilde \Omega$. For the ground state vibrations, the
renormalization is given by the Franck-Condon factor $M_{0,0} = \exp[- V_0^2 / (2 \hbar ^2
\omega_{ph}^2)]$. Thus, the effect of the vibrations is to slow down the electron
dynamics. All control strategies remain unchanged, but the time for achieving an initial
localization increases to $\tilde \tau_L = (\pi/ 2 \Omega) \exp[ V_0^2 / (2 \hbar ^2
\omega_{ph}^2)] $. For an application of this model to the DW coupled to localized
phonons, we may choose the Debye temperature as the maximum phonon energy, i.e. $\hbar
\omega_{ph} = k_B \Theta$ \cite{adachi1, adachi2}, and measure the coupling constant $V_0$
in units of $\hbar \omega_{ph}$. In the phonon-free system, the characteristic time is
$T_0 \approx 80$~fs. Due to electron-phonon ($e\mhyphen ph$) coupling, it increases to
$\tilde T_0 \approx 110$~fs.

\paragraph*{(ii) Localized coupling $v_c(x) = \exp[-(x-x_c)^2/2 \lambda^2]$.}  Substituting
again $x \rightarrow -x_0 \sigma_x/2$, and evaluating the matrix exponent yields
\vspace{-0.1cm}
\begin{equation}
H_C \rightarrow V_0 \frac{y}{y_0} \exp\left[-\frac{x_0^2/4 + x_c^2}{2 \lambda^2}\right]
(\cosh \xi - \sigma_x \sinh \xi), \vspace{-0.1cm}
\end{equation}
with $\xi = x_0 x_c/2\lambda^2$. If the coupling to the phonon degree of freedom is
centered around the potential barrier, i.e. $x_c = 0$ and therefore $\sinh \xi = 0$, $H_C$
will no longer depend on the electron term. The two degrees of freedom are then
decoupled. For the full spectrum system, this means that the phonon effect plays a
minor role because the wave function nearly vanishes at $x=0$.

 The worst case scenario in terms of invalidating previous localization schemes is the
 $e\mhyphen ph$ interaction acting in one of the quantum wells, since the resulting
 problem has now the maximal asymmetry. For the TLS model (after the polaron
 transformation), this asymmetry leads to an additional term $H_{TLS} = -\hbar \tilde
 \Omega \sigma_z / 2 \mp \hbar \kappa \sigma_x/2$ for $x_c = \mp x_0/2$. The parameter
 $\kappa$ is quadratic in $V_0$: $\kappa = q(\lambda) V_0^2/\hbar^2 \omega_{ph}$, where
 $q(\lambda) = 1- e^{-x_0^2/2 \lambda^2}$ is a prefactor describing the spatial extension
 of the $e\mhyphen ph$ interaction along the structure.  For this situation we inferred
 the following results:\\ (a) \textit{ Localizing the electron in the left well}: Being no
 longer an eigenstate of the uncoupled Hamiltonian, the symmetric ground state of the DW
 potential oscillates and therefore gains the momentum $\mp \delta p_1$ for $x_c=\mp
 x_0/2$ and $t<T_0/4$. The corresponding pulse angle $\alpha$ must be modified according
 to $\alpha = \pi/2 \mp \Delta \beta_1$ with $\Delta \beta_1 = 2 \mu \delta p_1 /
 \hbar$.\\ (b) \emph{Maintaining the localization}: Due to the asymmetry, the electron
 tunnels back from the left well with an increased (a decreased) momentum $p \pm \delta
 p_2$ for $x_c=x_0/2$ ($x_c=-x_0/2$). Note that this additional momentum $\delta p_2 $
 depends on the time. In order to incorporate the dynamics arising from $\delta p_2$, the
 strategy is now to use the same pulse sequence, but to adjust the pulse angle to $\alpha
 = \pi \mp \Delta \beta_2$ with $\Delta \beta_2(t) = 2 \mu \delta p_2(t) / \hbar$. \\ The
 angles $\Delta \beta_1$ and $\Delta \beta_2$, or, equivalently the transferred momenta,
 can be calculated using the TLS model.\\ Alternatively, one can also compensate the
 asymmetric vibrational coupling by a constant electric field $E_0$ oriented in the
 $x$-direction.  In the presence of the electric field $E^{(\pm)}_0 = \mp \hbar \kappa/2
 \mu$, the TLS term proportional to $\sigma_x$ is exactly canceled. Thus, if the value
 $E^{(\pm)}_0$ is tuned appropriately, the control strategies remain the same as in the
 case of zero coupling.

In order to compare the optimal values obtained from the TLS model with the exact results,
we performed a full numerical propagation based on the two-dimensional split-operator
approach~\cite{splitop}. For definiteness we consider $x_c = -x_0 / 2$, the case $x_c =
x_0 / 2$ can be treated analogously. For clarity we introduce the dimensionless coupling
parameter $\gamma = 2 V_0 / \hbar \omega_{ph}$. Our control quantity is the probability of
finding the electron in the left well $P_L(t)$.
\paragraph{Localizing the electron in the left well}

\begin{figure}
	\centering
		\includegraphics[width=\columnwidth]{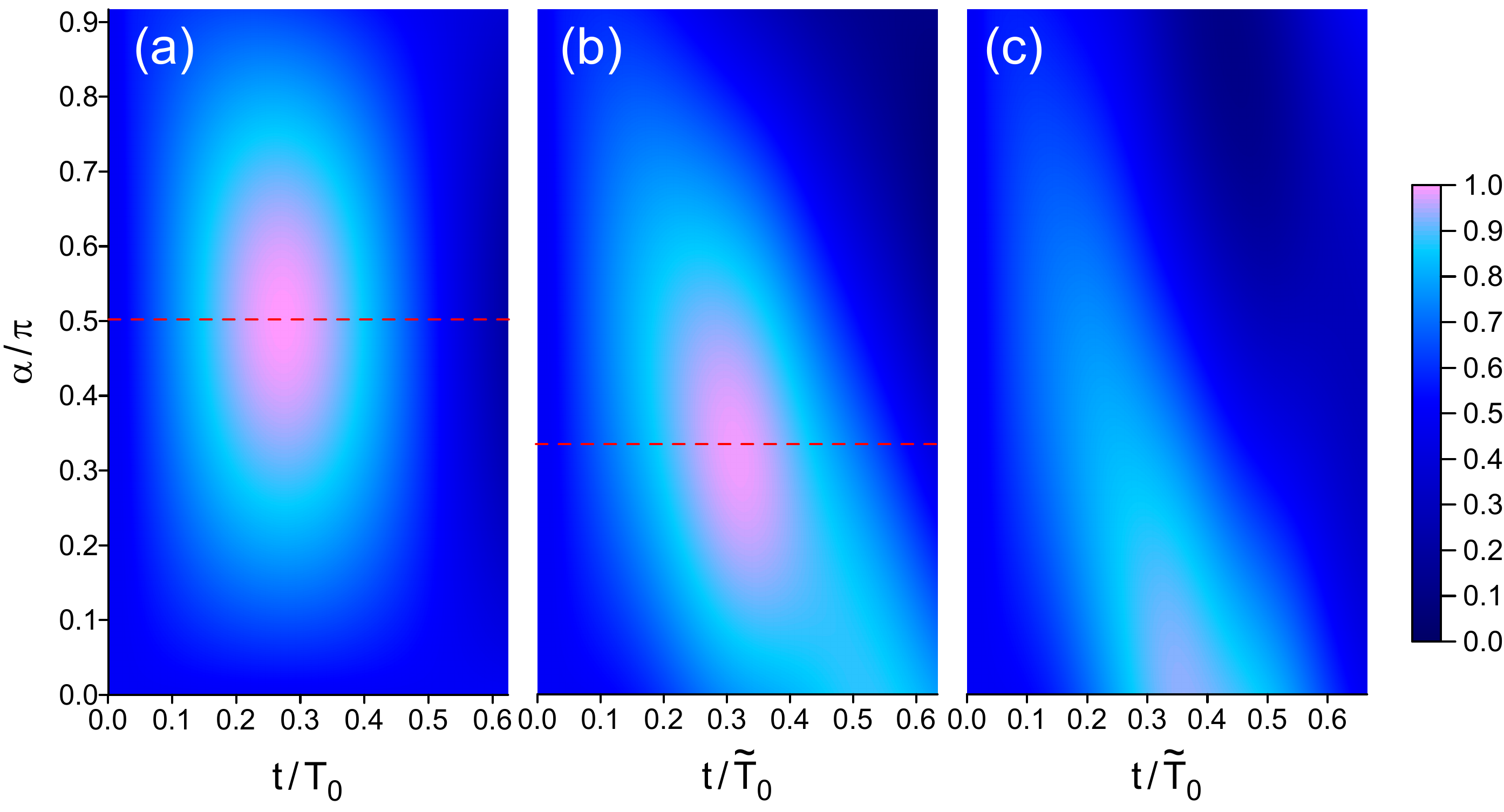}\vspace{-0.2cm}
		\caption{The time-dependent probability $P_L$ as a function of the applied
		laser pulse angle $\alpha=\frac{2 \mu \Delta p}{\hbar}$ for different
		values of the rescaled coupling (a) $\gamma := 2 V_0 / \hbar \omega_{ph} =
		0$, (b) $\gamma = 0.6$, and (c) $\gamma = 1$.}\vspace{-0.3cm}
		\label{plot2}
\end{figure}

\begin{figure}[H]
    \centering
		\vspace{-0.2cm}\includegraphics[width=7cm]{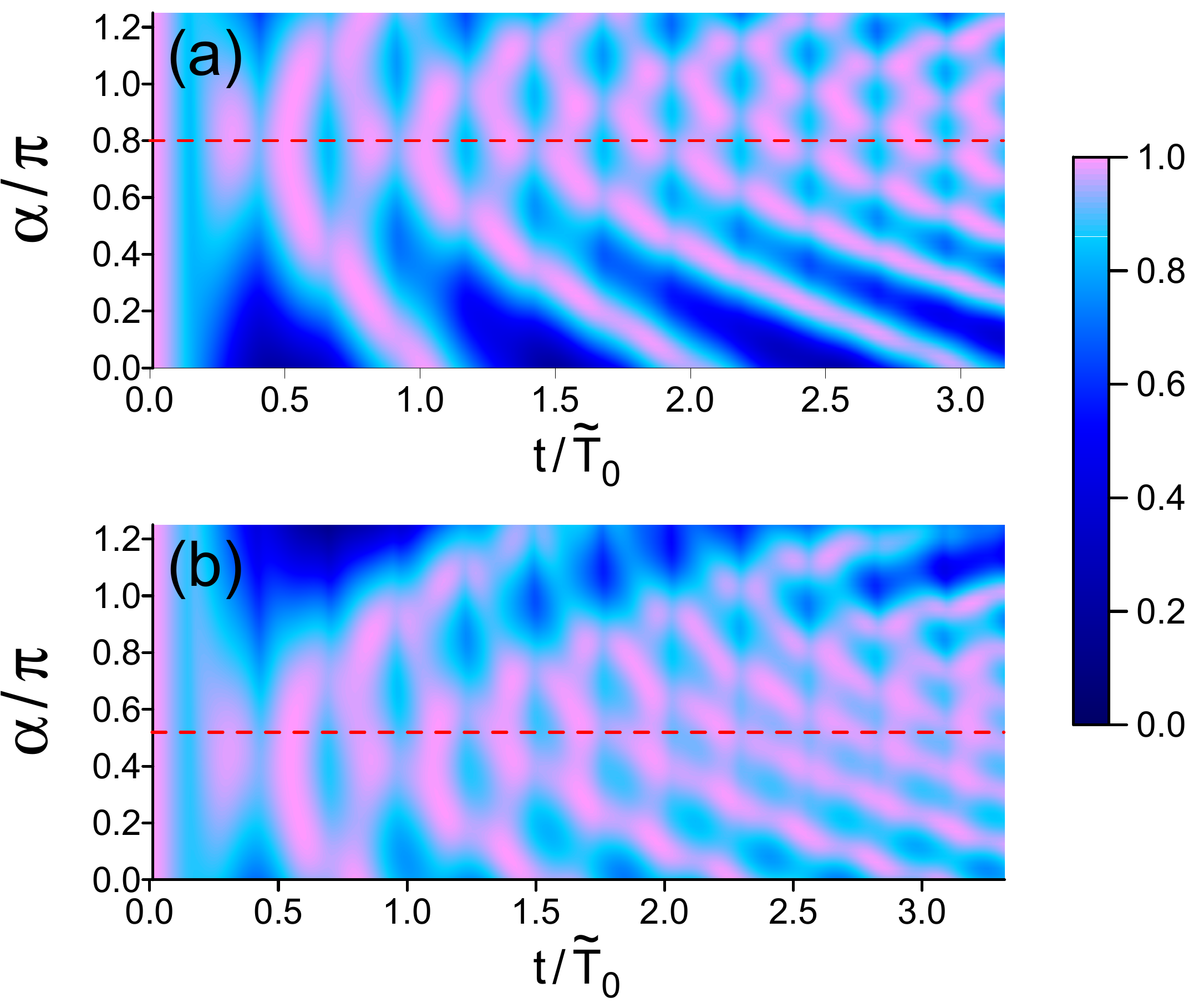}\vspace{-0.2cm}
		\caption{The time-dependent probability $P_L$ as a function of the pulse
		angle $\alpha$ of the train of laser pulses with the $T = 80$~fs for
		values of the coupling parameter: (a) $\gamma = 0.6$, and (b) $\gamma = 1$.}
		\label{plot4}
\end{figure}

In fig. \eqref{plot2}, the time scale is given by the renormalized characteristic time
$\tilde T_0$. The red dashed lines denote the optimal pulse angles obtained from the TLS
model, which agrees well with the numerical result. In the case of full coupling
fig.~(\ref{plot2}~c), an initial pulse
\begin{figure}[H]
	\centering
		\includegraphics[width=\columnwidth]{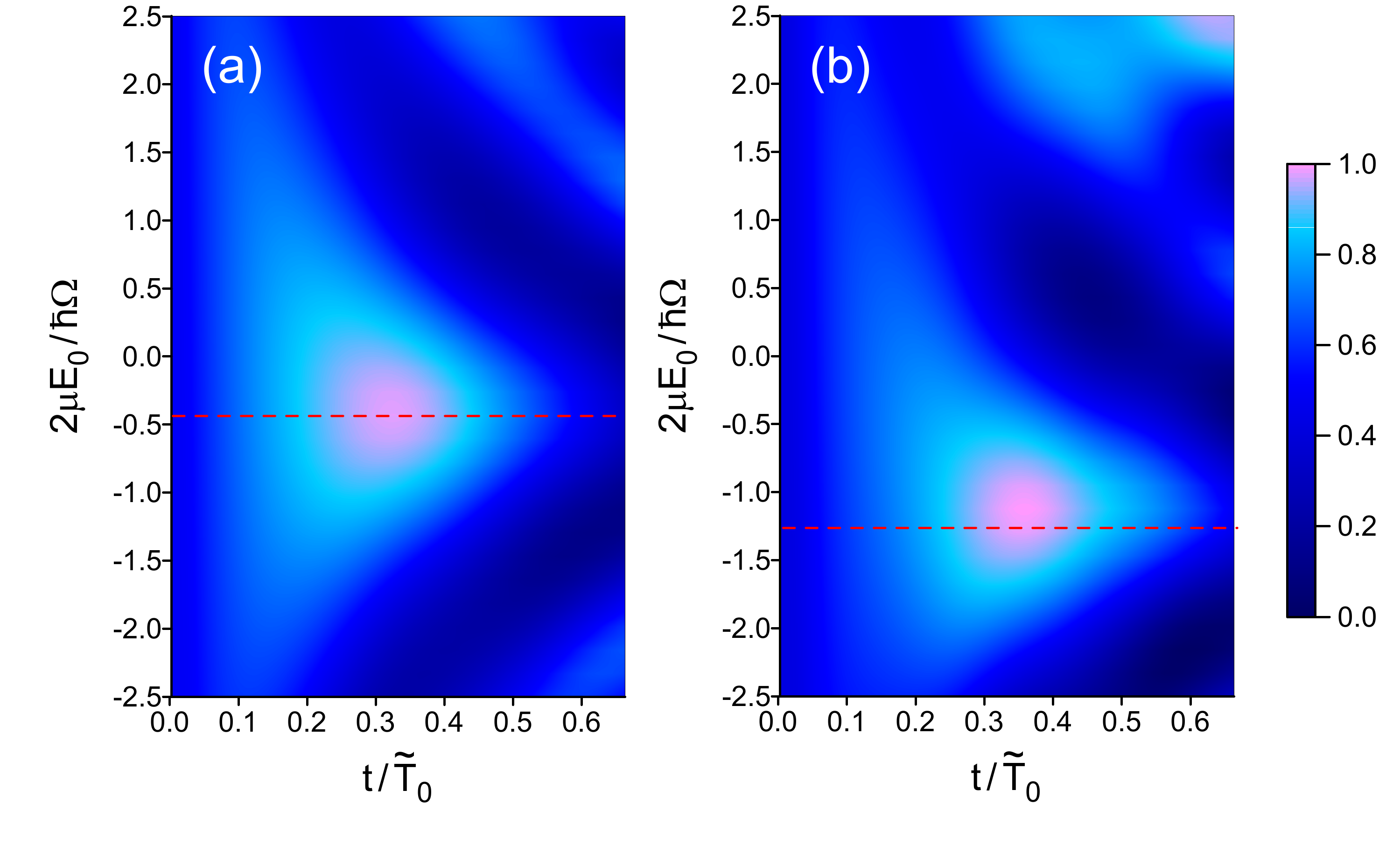}\vspace{-0.1cm}
		\caption{The time-dependent probability $P_L$, if additionally to the
		localizing laser pulse with $\alpha = \pi/2$ a constant electric field
		$E_0$ (measured in units of $2 \mu /\hbar \Omega$) is applied, for
		different values of the coupling parameter: (a) $\gamma = 0.6$, and (b)
		$\gamma =1$. \vspace{-0.3cm}}
		\label{plot1}
\end{figure}
with any amplitude can not totally localize the
electron, but $P_L$ still reaches values around 90\%. In order to obtain full control over
the electron in this scenario as well, one can use a constant electric field, as explained
above. We apply the unmodified laser pulse with a pulse angle $\alpha =
\pi/2$. Fig. \eqref{plot1} shows the effect on the dynamical localization.
The constant electric field $E_0$ in fig. \eqref{plot1} is rescaled in units of $2 \mu /
\hbar \Omega$ so as to compare with  $\kappa / \Omega$
(red dashed lines).\vspace{-0.05cm}
\paragraph{Maintaining the localization}
If the electron has now been transferred predominantly to the left well, a periodic train
 of pulses with a pulse angle $\alpha = \pi - \Delta \beta_2$ for zero coupling leads to a
 very high degree of localization. Fig. (\ref{plot4}) shows again the probability $P_L(t)$ but with
 the initial condition $P_L(t=0) = 1$. Since the period of the pulses has been chosen as
 $T = 80$~fs, $P_L$ shows a maximum for multiples of $T$, even in the case of a
 vibrational coupling. The red dashed lines represents the
 TLS results. Note that in fig. (\ref{plot4}), the averaged localization is over 90 \%.
\paragraph*{Conclusions}
We analyzed  analytically and numerically  the electron dynamics in a DW
coupled to vibrational modes.  Mapping onto a spin-model and using the polaron transformation we found
analytical expressions for i) times and ii) light pulse parameters that are appropriate
for switching the system from the initially delocalized to the localized target state and
to maintain the localization. These estimates are in excellent agreement with the
full-fledged two-dimensional numerical propagation. We considered two cases of uniform and
localized vibrational coupling, and also when a static electric field is
applied. Even in the presence of the vibrational environment we achieve an averaged
localization of 90\% which makes this system a suitable candidate for applications in
quantum information or as an ultra-fast optical switch.

\vspace{-0.3cm}
\small{
}


\begin{thebibliography}{17}
\bibitem{remark} For numerical simulations we used HCP of $\sin^2(\pi t/\tau_p)$ shape and
  $\tau_p$ duration. Additional studies were made to clarify the role of tails of
  realistic HCPs on the electron dynamics. They confirm that our control scenarios are
  robust with respect to these effects.
\end{thebibliography}
\end{document}